\documentclass[twocolumn]{aastex63}

\received{January 1, 2099}
\revised{January 1, 2099}
\accepted{January 1, 2099}

\shorttitle{Intrusion of Cosmic Rays}
\shortauthors{Fujita et al.}

 \usepackage[T1]{fontenc}

\begin{document}

\title{Intrusion of MeV-TeV Cosmic Rays into Molecular Clouds Studied by Ionization, the Neutral Iron Line, and Gamma Rays}

\correspondingauthor{Yutaka Fujita}
\email{y-fujita@tmu.ac.jp}

\author[0000-0003-0058-9719]{Yutaka Fujita}
\affiliation{Department of Physics, Graduate School of Science,\\
Tokyo Metropolitan University, 1-1 Minami-Osawa,\\
Hachioji-shi, Tokyo 192-0397, Japan}

\author{Kumiko K. Nobukawa}
\affiliation{Faculty of Science and Engineering, Kindai University,\\ 3-4-1 Kowakae, Higashi-Osaka, 577-8502, Japan}

\author{Hidetoshi Sano}
\affiliation{National Astronomical Observatory of Japan, Mitaka, Tokyo 181-8588, Japan}


\begin{abstract}
Low-energy ($\sim$~MeV) cosmic rays (CRs) ionize molecular clouds and
create the neutral iron line (Fe {\small I} K$\alpha$) at 6.4~keV. On
the other hand, high-energy ($\gtrsim$~GeV) CRs interact with the dense
cloud gas and produce gamma rays. Based on a one-dimensional model, we
study the spatial correlation among ionization rates of gas, 6.4~keV
line fluxes, and gamma-ray emissions from a molecular cloud illuminated
by CRs accelerated at an adjacent supernova remnant. We find that the
spatial distributions of these three observables depend on how CRs
intrude the cloud and on the internal structure of the cloud. If the
intrusion is represented by slow diffusion, the 6.4~keV line should be
detected around the cloud edge where ionization rates are high. On the
other hand, if CRs freely stream in the cloud, the 6.4~keV line should
be observed where gamma rays are emitted.  In the former, the cooling
time of the CRs responsible for the 6.4~keV line is shorter than their
cloud crossing time, and it is opposite in the latter. Although
we compare the results with observations, we cannot conclude whether the
diffusion or the free-streaming is dominantly realized. Our predictions
can be checked in more detail with future X-ray missions such as 
XRISM and Athena and by observations of ionization rates that
cover wider fields.
\end{abstract}


\section{Introduction} 
\label{sec:intro}

Supernova remnants (SNRs) are considered a main source of cosmic rays
(CRs\footnote{We consider protons as CRs in this paper.}) in the
Milky way. Some SNRs are surrounded by molecular clouds and are even
interacting with them. Gamma-ray emissions have been observed from such
molecular clouds
\citep[e.g.][]{2007ApJ...664L..87A,2008A&A...481..401A,2009ApJ...698L.133A,2010Sci...327.1103A,2010ApJ...710L.151T,2013Sci...339..807A},
which probably means that CRs have escaped from the SNRs and are
producing the gamma rays through interaction with the dense gas.
However, the gamma-ray observations do not detect all the CRs
accelerated at the SNRs.  While the gamma rays are created via
$pp$-interaction, only CRs with energies of $E\gtrsim$~GeV can exceed
the threshold of the interaction. This means that CRs with $E\sim$~MeV
cannot be observed in gamma rays.

MeV CRs are expected to be around SNRs because they should have been
accelerated at the SNR shocks as with the gamma-ray emitting CRs with
higher energies. Their existence has been indicated by previous studies.
First, the ionization rates of dense clouds around SNRs are higher than
those in the general Galactic interstellar medium
\citep{2010ApJ...724.1357I,2011ApJ...740L...4C,2014A&A...568A..50V}.
Second, the 6.4~keV neutral iron line (Fe {\small I} K$\alpha$) has been
detected for several SNRs
\citep{2014PASJ...66..124S,2016PASJ...68S...8S,2018ApJ...854...87N,2018PASJ...70...35O,2018ApJ...854...71B,2018PASJ...70...23S,2019PASJ...71..115N}. Both
the ionization and the 6.4~keV line can be attributed to the interaction
of MeV CRs with dense gas.

Recently, \citet{2019PASJ...71...78M} and \citet{2019PASJ...71..115N}
showed that both the 6.4~keV line flux and gamma rays from a few SNRs
can be explained by the CRs that were accelerated at the SNRs.
\citet{2020A&A...635A..40P} indicated that the MeV CRs that ionize dense
gas around the SNR W~28 belong to the same CR population that is also
generating gamma-ray emissions. These studies suggest that we can
discuss broad CR spectra from MeV to TeV by combining the three
observables (6.4~keV line fluxes, ionization rates, and gamma rays).

Based on this fact, we show in this paper that the three observables can
be used to study the propagation of CRs in molecular clouds. Some
previous studies \citep[e.g.][]{2015MNRAS.451L.100M,2018MNRAS.480.5167P}
assumed that CRs stream freely inside clouds because Alfv\'en waves that
scatter CRs are dumped due to ion-neutral friction
\citep{1982ApJ...259..859Z}. Using numerical simulations, on the other
hand, \citet{2019ApJ...872...46I} showed that CR streaming generates
Alfv\'en waves and makes the CR propagation diffusive \citep[see
also][]{2018ApJ...855...23I,2018ApJ...868..114D,2019ApJ...879...14S}. Since CR propagation
in dense clouds depends on complicated microphysics, observational work
is an important means to test and anchor theoretical predictions on
which type of propagation (free-streaming or diffusion) is being
realized. We show that this issue can be addressed using the above three
observables. We focus on a molecular cloud illuminated by CRs
accelerated at an adjacent SNR.

This paper is organized as follows. In section~\ref{sec:model}, we
describe our models for CR propagation and ionization in a dense cloud,
and emissions from the cloud. In section~\ref{sec:resu}, we discuss the
profiles of the 6.4~keV line flux, ionization rates, and gamma-ray
emissions from the cloud and argue their dependence on the CR propagation
models. In section~\ref{sec:disc}, we compare our predictions with
observations. The conclusion of this paper is presented in
section~\ref{sec:conc}.

\section{Models}
\label{sec:model}

\subsection{CR Propagation}
\label{sec:diff}

We consider two possibilities for the propagation of
CRs: a diffusive case and a free-streaming
case. 

\subsubsection{Diffusive Case}

If MHD turbulence and/or waves are developed in molecular clouds, CR
prorogation may be diffusive. In this case, we can study the intrusion
of CRs into clouds by solving a one-dimensional (1D) diffusion-advection
equation.
\begin{equation}
\label{eq:diffadv}
 \frac{\partial f}{\partial t} = \frac{\partial}{\partial x}\left(D\frac{\partial f}{\partial x}\right)
- v\frac{\partial f}{\partial x} - \frac{1}{p^2}\frac{\partial}{\partial p}(\dot{p}p^2 f)
\:,
\end{equation}
where $f=f(t,x,p)$ is the CR particle distribution function, $t$ is the
time, $x$ is the inward distance from the surface of the molecular
cloud, $v$ is the bulk velocity of the CR particles, $p$ is the particle
momentum, $D=D(p)$ is the diffusion coefficient, and $\dot{p}$ is the
rate of momentum loss of CRs due to interaction with gas. The diffusion
coefficient is assumed to be a simple power-law form and is defined
based on a standard value in the Milky way:
\begin{equation}
\label{eq:Dp}
 D(p) = 1\times 10^{28}\chi\frac{v_p}{c}
\left(\frac{pc}{\rm 10~GeV}\right)^{\delta}
\left(\frac{B}{3\:\rm \mu G}\right)^{-\delta} \:\rm cm^2\: s^{-1}\:,
\end{equation}
where $v_p$ is the particle velocity corresponding to a momentum $p$,
$c$ is the light speed, and $B$ is the magnetic field
\citep[e.g.][]{2009MNRAS.396.1629G}. We assume Kolmogorov type
turbulence ($\delta=1/3$), unless otherwise mentioned. We introduce the
reduction factor $\chi (<1)$ because the coefficient around SNRs can be
reduced by waves and/or turbulence generated through the stream of
escaping CRs
\citep{1969ApJ...156..445K,1969ApJ...156..303W,2008AdSpR..42..486P,2010ApJ...712L.153F,2011MNRAS.415.3434F}. For
the momentum loss of CRs ($\dot{p}$), we consider ionization loss and
pion production \citep{1994A&A...286..983M}.  The cooling is
particularly effective for CRs with lower energies (see Figure~2 in
\citealt{2018MNRAS.480.5167P}). The edges of the cloud are at $x=0$ and
$x_{\rm max}$, and we solve equation~(\ref{eq:diffadv}) at $0<x<x_{\rm
max}$. CRs are injected at $x=0$. For the sake of simplicity, we assume
that CRs are confined in the cloud and thus we adopt a reflective
boundary condition, $\partial f/\partial x=0$, at $x=x_{\rm max}$. The
actual boundary condition may depend on the CRs and/or magnetic fields
outside the cloud ($x\geq x_{\rm max}$), which are beyond the scope of
this paper. We take the advection velocity of $v=300\rm\: km\: s^{-1}$,
which was used as an expansion velocity of SNRs in
\citet{2019PASJ...71...78M}. The results are not sensitive to $v$, as
long as it is small enough.

\subsubsection{Free-Streaming Case}

If MHD turbulence and waves are damped in molecular clouds, CRs may
freely stream along magnetic field lines. For the sake of simplicity, we
assume that magnetic fields are parallel to the $x$-direction and CRs
move along them with the speed of $v_p$. If CRs are continuously
injected at $x=0$ for $t>0$ and if the cooling can be ignored, CRs with
a momentum $p$ reach the boundary $x=x_{\rm max}$ at $t\sim x_{\rm
max}/v_p$. Moreover, if a reflective boundary is assumed at $x=x_{\rm
max}$, the CRs are almost uniformly distributed in the cloud at
$t\gtrsim x_{\rm max}/v_p$. This evolution can be mimicked using
equation~(\ref{eq:diffadv}) by setting the diffusion coefficient as
\begin{equation}
 D(x,p) = v_p x/2\:.
\end{equation}
This is because while CRs injected at $t=0$ should reach $x\sim v_p t$
at time $t$, the 1D diffusion velocity at $x$ is given by
\begin{equation}
 v_{\rm diff} = \frac{\sqrt{2 D(x,p) t}}{t}
 = \frac{\sqrt{v_p x t}}{t} \sim v_p
\end{equation}
for $t\lesssim x_{\rm max}/v_p$. The advantage of this formulation is
that the CR distribution at $t\lesssim x_{\rm max}/v_p$ is smoothly
transformed into that at $t\gtrsim x_{\rm max}/v_p$.

\subsection{Molecular Cloud and CR Injection}

We construct a model of a molecular cloud and CR injection into it so
that the results of calculations are roughly consistent with the
observations of the northeastern cloud close to the SNR W~28. The age of
W~28 is $\sim 3 \times 10^4$--$4\times 10^4$~yr
\citep{2002ApJ...575..201R,2002AJ....124.2145V,2018ApJ...860...69C}. We
emphasize that the model may not be the only solution to explain the
observations for W~28 considering the simplicity of the model and the
uncertainties of observations.

\begin{figure}[t]
\plotone{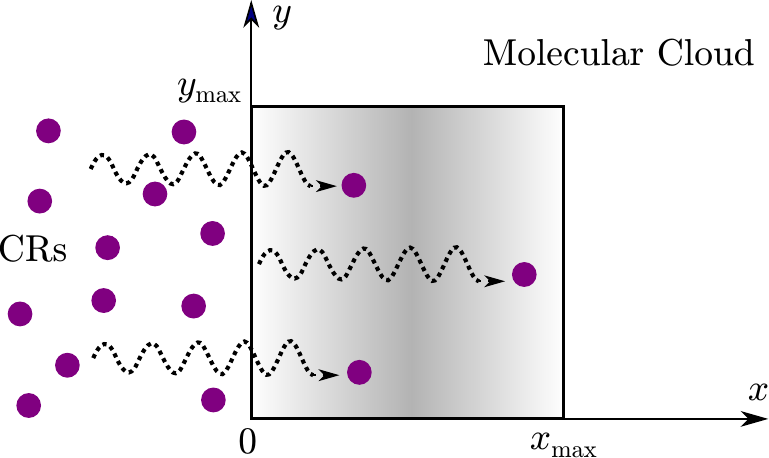} \caption{Schematic figure of CRs and a molecular
cloud. CRs at $x<0$ intrude the molecular cloud at $0<x<x_{\rm
max}$. The gas density of the cloud reaches the maximum at $x=x_{\rm
max}/2$ as is shown by the gray shading. \label{fig:mol}}
\end{figure}

We assume that the cloud is a cube 8~pc on each side, and it is in
contact with an SNR with a volume of $V_{\rm SNR} = 4\pi (10~{\rm
pc})^3/3$. The directions of the sides are defined as $x$, $y$, and $z$
(Figure~\ref{fig:mol}), although we solve equation~(\ref{eq:diffadv})
only in the $x$-direction ($0<x<x_{\rm max}=8$~pc). We assume that
$x_{\rm max}=y_{\rm max}=z_{\rm max}$. The hydrogen number density is
given by
\begin{equation}
 \label{eq:nH}
 n_{\rm H} = n_{\rm H0}\exp\left[-\frac{(x-x_0)^2}{x_{\rm core}^2}\right]
\:.
\end{equation}
We assume that $x_0=x_{\rm max}/2=4$~pc, $x_{\rm core}=2$~pc, and
$n_{\rm H0}=1\times 10^4\rm\: cm^{-3}$, which gives the total cloud mass
of $M_{\rm mol}=5.6\times 10^4\: M_\sun$. This mass is close to the
observed one ($M_{\rm mol}\sim 5\times 10^4\: M_\sun$;
\citealt{2008A&A...481..401A}). At $x=0$ and $x=x_{\rm max}$, the
density is $n_{\rm H}=183\rm\: cm^{-3}$.

At $t=0$, the cloud does not contain CRs. At $t>0$, CRs are injected at
$x=0$. The CR spectrum at $x=0$ is 
represented as a broken power law:
\begin{equation}
\label{eq:f0}
  f_0(p) = \left\{ \begin{array}{ll}
    C\left(\frac{pc}{\rm GeV}\right)^{-\alpha} & (pc>1~{\rm GeV}) \\
    C\left(\frac{pc}{\rm GeV}\right)^{-2} & (pc<1~{\rm GeV})
  \end{array} \right.\:,
\vspace{2mm}
\end{equation}
where $\alpha~(>2)$ is a parameter. The normalization $C$ is given so
that the total energy density of CRs at $pc>1$~GeV is $E_{\rm
CR,tot}/V_{\rm SNR}$, where $E_{\rm CR,tot}$ is the total energy of CRs
with $pc>1$~GeV accelerated at the SNR. The reason we adopt the broken
power-law form is that if we assume a single power-law form with the
index of $\alpha$, overabundant low-energy CRs ionize the cloud
excessively and produce an overly bright 6.4~keV line for a given
gamma-ray luminosity. This suggests that the low-energy CRs are injected
somewhat differently from the higher-energy CRs. In the scenario
proposed by \citet{2019PASJ...71...78M}, the higher-energy CRs ($\gtrsim
1$~GeV) once escaped from the SNR when it was much younger and
smaller. Then they diffuse in the interstellar space and enter the
cloud. On the other hand, lower-energy CRs ($\lesssim 1$~GeV) are
directly injected into the cloud from the SNR. This steepens the CR
spectrum at $\gtrsim 1$~GeV because the CRs in the interstellar space are
affected by energy-dependent diffusion in such a way that CRs with
higher energies diffuse faster. Although the momentum-dependent escape
process of the CRs must be considered in order to precisely derive the
form of $f_0(p)$
\citep[e.g.][]{2010MNRAS.409L..35L,2011MNRAS.410.1577O}, we here adopt
the simplified model (equation~(\ref{eq:f0})).  It is to be noted that
\citet{2020A&A...635A..40P} introduced a lower-energy limit in their CR
spectrum instead of a broken power law.

We assume that the injection (equation~(\ref{eq:f0})) is
time-independent at $t>0$ for the sake of simplicity.  We consider the
intrusion of CRs on time-scales of $\leq 1\times 10^4$~yr, which is much
smaller than the age of the SNR ($\sim 3 \times 10^4$--$4\times
10^4$~yr). Thus, the injection may not change much during that period
after the contact with the SNR.

\subsection{Ionization Rates and Emissions}

We consider the ionization of H$_2$, and the 6.4~keV line and gamma-ray
emissions by CR protons. We ignore CR electrons because their
contribution is subdominant for the 6.4~keV line and gamma-ray emissions
for molecular clouds around SNRs
\citep{2018ApJ...854...87N,2020A&A...635A..40P}. The ionization of
molecular clouds can also be explained by protons
\citep{2020A&A...635A..40P}.

The ionization rate is given by
\begin{eqnarray}
 \zeta_p({\rm H}_2) &=& 
\int_{I({\rm H}_2)}^{E_{\rm max}}n_{\rm CR}(E)v_p[1 + \phi_p(E)]\sigma^p_{\rm ion}(E)dE
\nonumber\\
& & 
+ \int_0^{E_{\rm max}}n_{\rm CR}(E)v_p\sigma_{\rm ec}(E)dE\:,
\label{eq:zetap}
\end{eqnarray}
where $E$ is the kinetic energy of a particle corresponding to
a momentum $p$, which means that $E=\sqrt{p^2 c^2 + m_p^2 c^4} - m_p
c^2$, where $m_p$ is the proton mass.  For given $t$ and $x$, the CR
density has a relation of $n_{\rm CR}(E)dE=4\pi p^2 f(t,x,p)dp$. In
equation~(\ref{eq:zetap}), $I({\rm H}_2) = 15.603$~eV is the ionization
potential of H$_2$, $E_{\rm max}~(=300~\rm TeV)$ is the maximum energy
considered, and $\phi_p(E)$ is a correction factor accounting for the
ionization of H$_2$ by secondary electrons
\citep{2009A&A...501..619P,2018MNRAS.480.5167P}.  For $\phi_p(E)$, we
adopted the one derived by \citet{2015ICRC...34..518K}. The quantities
$\sigma^p_{\rm ion}$ and $\sigma_{\rm ec}$ are the proton ionization
cross-section and the electron capture cross-section, respectively, and
we use those obtained by
\citet{1983PhRvA..28.3244R} and \citet{1985RvMP...57..965R}.

The photon number intensity of the 6.4~keV neutral iron line is given by
\begin{equation}
\label{eq:line formula}
I_{{\rm 6.4keV}}= \frac{1}{4\pi}\int dE\: \sigma_{\rm 6.4keV}(E) 
v_p \int d\ell \: n_{\rm H}(\ell) n_{{\rm CR}}(E,\ell) \:,
\end{equation}
where $\sigma_{\rm 6.4keV}(E)$, $n_{\rm H}(\ell)$, and $\ell$ are the
cross-section to produce the iron line at 6.4~keV, the number density of
hydrogens in the molecular cloud, and the coordinate along the line of
sight, respectively. Since the direction of $\ell$ is not
necessarily perpendicular to that of $x$, the CR density $n_{{\rm CR}}$
includes the dependence on $\ell$ in equation~(\ref{eq:line formula}).
We derive gamma-ray spectra using the models by
\citet{2006ApJ...647..692K}, \citet{2006PhRvD..74c4018K}, and
\citet{2008ApJ...674..278K} based on $n_{\rm H}(\ell)$ and $n_{\rm
CR}(E,\ell)$.

\section{Results}
\label{sec:resu}

We solve equation~(\ref{eq:diffadv}) using a standard implicit
method. We adopt 200 unequally spaced meshes in the $x$-direction that
cover the cloud.  The mesh at $x\sim 0$ has a width of $\sim 3\times
10^{-4}$~pc, and the width of the mesh at $x\sim x_{\rm max}$ is $\sim
0.3$~pc. In this section, we show the results of our calculations for
both the diffusive and free-streaming cases. We do not intend to
precisely reproduce observations, but we rather focus on how results
vary when parameters are changed. In particular, we study the
profiles of $\zeta_p({\rm H}_2)$, $I_{{\rm 6.4keV}}$, and gamma-ray
surface brightness $I_\gamma$.

\subsection{Diffusive Case}
\label{sec:rdiff}

\begin{figure}[t]
\plotone{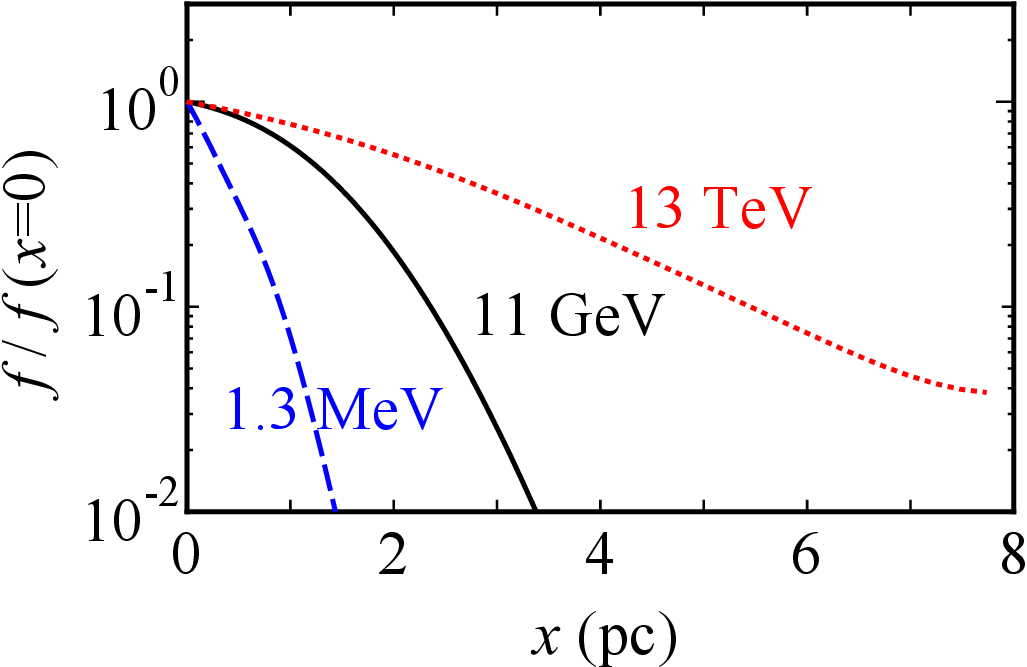} \caption{Distributions of CRs as functions of $x$ at
$t=3000$~yr for the diffusive case: $E=1.3$~MeV (blue dashed line),
$11$~GeV (black solid line), and 13~TeV (red dotted line)
\label{fig:dist_d}}
\end{figure}

\begin{figure}[t]
\plotone{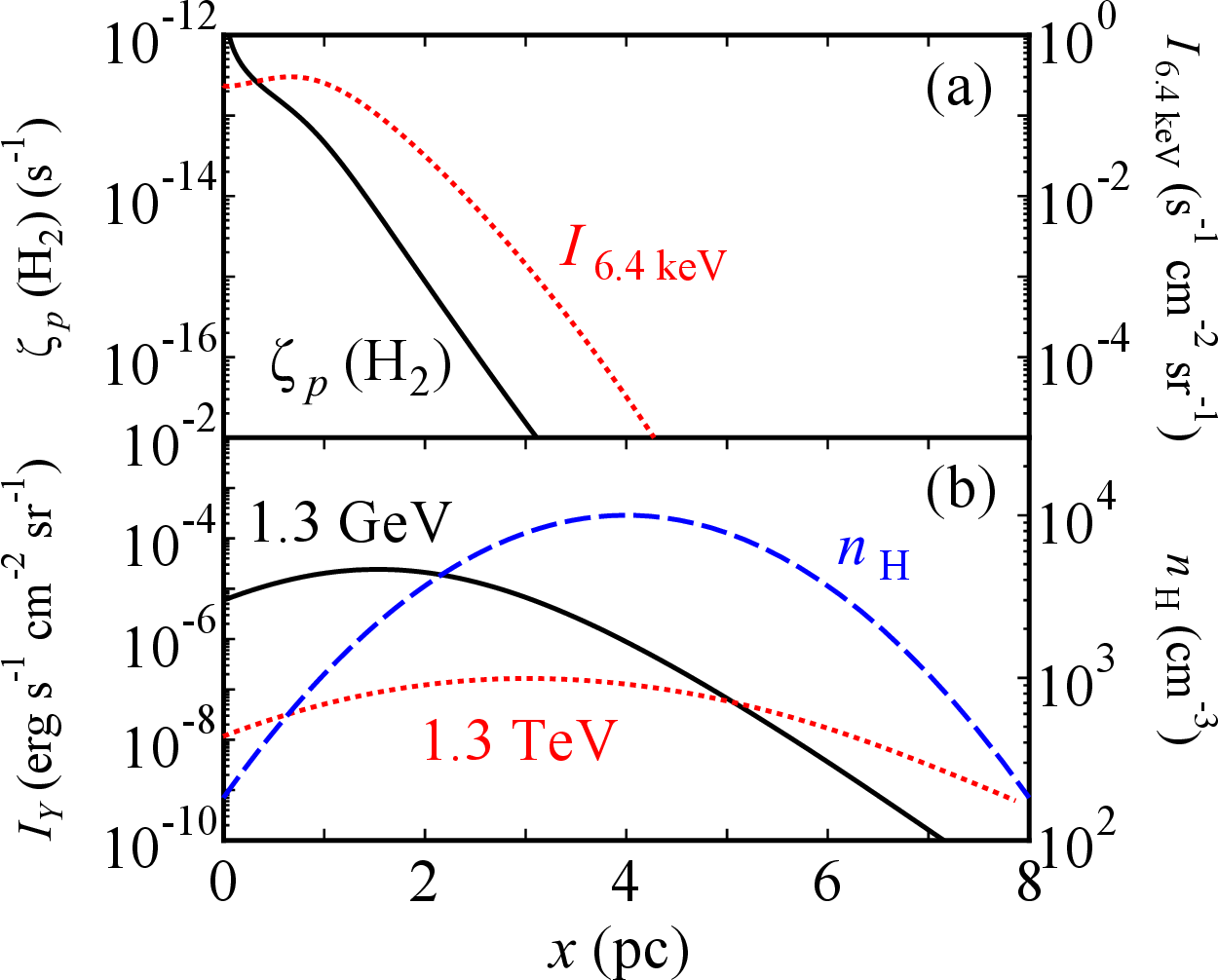} \caption{(a) Ionization rate ($\zeta_p({\rm H}_2)$;
black solid line) and surface brightness of the 6.4~keV line ($I_{{\rm
6.4keV}}$; red dotted line) as functions of $x$ at $t=3000$~yr for the
diffusive case. (b) Gamma-ray surface brightness ($I_\gamma$) profiles
at $E_\gamma=1.3$~GeV (black solid line) and $E_\gamma=1.3$~TeV (red
dotted line) at $t=3000$~yr for the diffusive case. Hydrogen density
profile (equation~(\ref{eq:nH})) is also shown (blue dashed
line). \label{fig:I_d}}
\end{figure}

Here, we assume that the line of sight ($\ell$) is parallel to the
$y$-axis and it is perpendicular to the $x$-axis (Figure~\ref{fig:mol}).
In equation~(\ref{eq:Dp}), we adopt a reduction factor of $\chi=0.01$,
which is a typical value around SNRs \citep{2009ApJ...707L.179F}. The
total energy of CRs with $pc>1$~GeV is $E_{\rm CR,tot}=1\times
10^{50}$~erg and the index is $\alpha=3.0$ (equation~(\ref{eq:f0})),
which are chosen so that the results are consistent with the observed
gamma-ray luminosity of W~28. Figure~\ref{fig:dist_d} shows the
distribution of CRs at $t=3000$~yr; CRs with higher energies can
penetrate deeper inside the cloud because the diffusion coefficient is
larger for them (equation~(\ref{eq:Dp})).  On the other hand,
lower-energy CRs ($\lesssim$~MeV) remain around $x\sim 0$ because they
are affected by rapid cooling as well as the smaller diffusion
coefficient. In Figure~\ref{fig:I_d}a, we present the profiles of the
ionization rate $\zeta_p({\rm H}_2)$ and the 6.4~keV line flux $I_{{\rm
6.4keV}}$. The former is a decreasing function of $x$ because
lower-energy CRs, which ionize the gas, are concentrated around $x\sim 0$
(Figure~\ref{fig:dist_d}). The 6.4~keV line emission is produced from
slightly deeper in the cloud compared with the region of large
$\zeta_p({\rm H}_2)$ (Figure~\ref{fig:I_d}a).  This is because while the
ionization is mainly caused by CR protons with $E\sim 0.01$~MeV
(Figure~1 in \citealt{2009A&A...501..619P}), the 6.4~keV line is
attributed to those with $E\sim 10$~MeV (Figure~5 in
\citealt{2012A&A...546A..88T}).  The latter CRs are less affected by
cooling and diffuse more inside the cloud. The line intensity $I_{{\rm
6.4keV}}$ has a peak at $x\sim 0.7$~pc, because it reflects not only the
CR density $n_{\rm CR}$ but also the hydrogen density $n_{\rm H}$
(equation~(\ref{eq:line formula}) and Figure~\ref{fig:I_d}b).  We note
that the large values of $\zeta_p({\rm H}_2)$ and $I_{{\rm 6.4keV}}$ at
$x\sim 0$ should be regarded as the upper limits because it is unlikely
that actual clouds are perfectly uniform along the line of sight.

Contrary to the large $\zeta_p({\rm H}_2)$ and $I_{{\rm 6.4keV}}$
regions that are limited to $x\sim 0$, gamma-ray emissions are produced
from almost the entire cloud especially for TeV gamma rays
(Figure~\ref{fig:I_d}b). This is because CRs with higher energies
($\gtrsim$~TeV) can penetrate the cloud (Figure~\ref{fig:dist_d}). Since
the gamma-ray emissivity is proportional to $n_{\rm H}$, the gamma-ray
profiles are subject to equation~(\ref{eq:nH}) that peaks at
$x=4$~pc. The gamma-ray fluxes from the whole cloud are
$F_\gamma=1.1\times 10^{-10}\rm\: erg\: cm^{-2}\: s^{-1}$ at
$E_\gamma=1.3$~GeV, and $F_\gamma=1.1\times 10^{-12}\rm\: erg\:
cm^{-2}\: s^{-1}$ at $E_\gamma=1.3$~TeV, which are close to the observed
values for W~28
\citep{2008A&A...481..401A,2010ApJ...718..348A,2018ApJ...860...69C}. It
is to be noted that CRs with an energy of $E$ produce gamma rays with an
energy of $E_\gamma \sim 0.1\: E$. In summary, the high $\zeta_p({\rm
H}_2)$ region should be observed close to the edge of the cloud, and
$I_{{\rm 6.4keV}}$ has a similar distribution. On the other hand, the
peak of the gamma-ray surface brightness is located deeper in the
cloud. In particular, the profile of TeV gamma rays rather follows that
of $n_{\rm H}$.

\begin{figure}[t]
\plotone{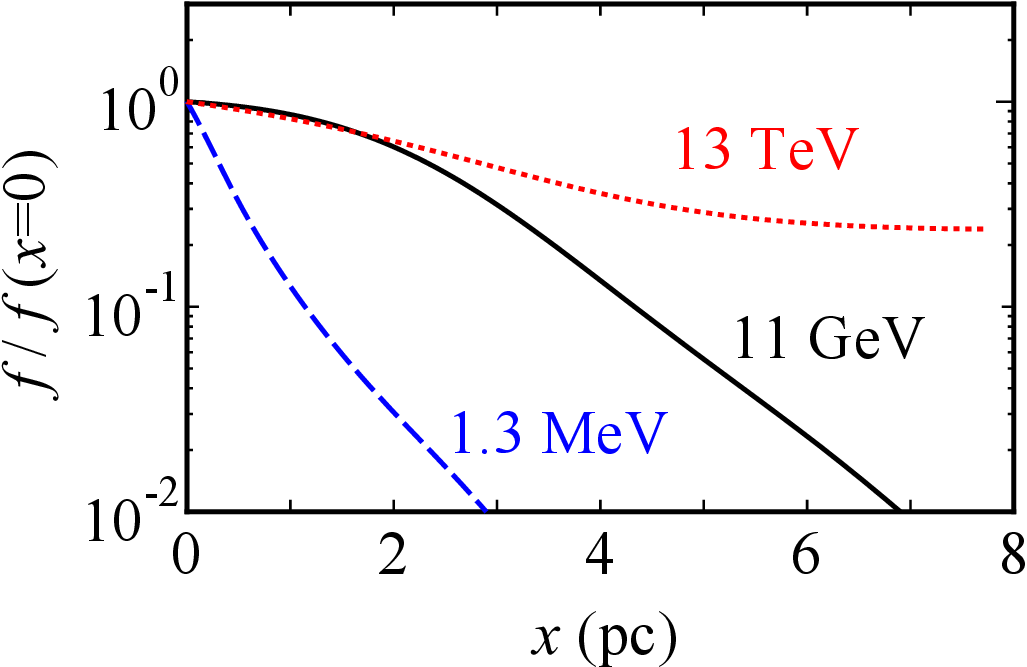} \caption{Same as Figure~\ref{fig:dist_d} but for $t=1\times 10^4$~yr. \label{fig:dist_d2}}
\end{figure}

\begin{figure}[t]
\vspace{2mm}
\plotone{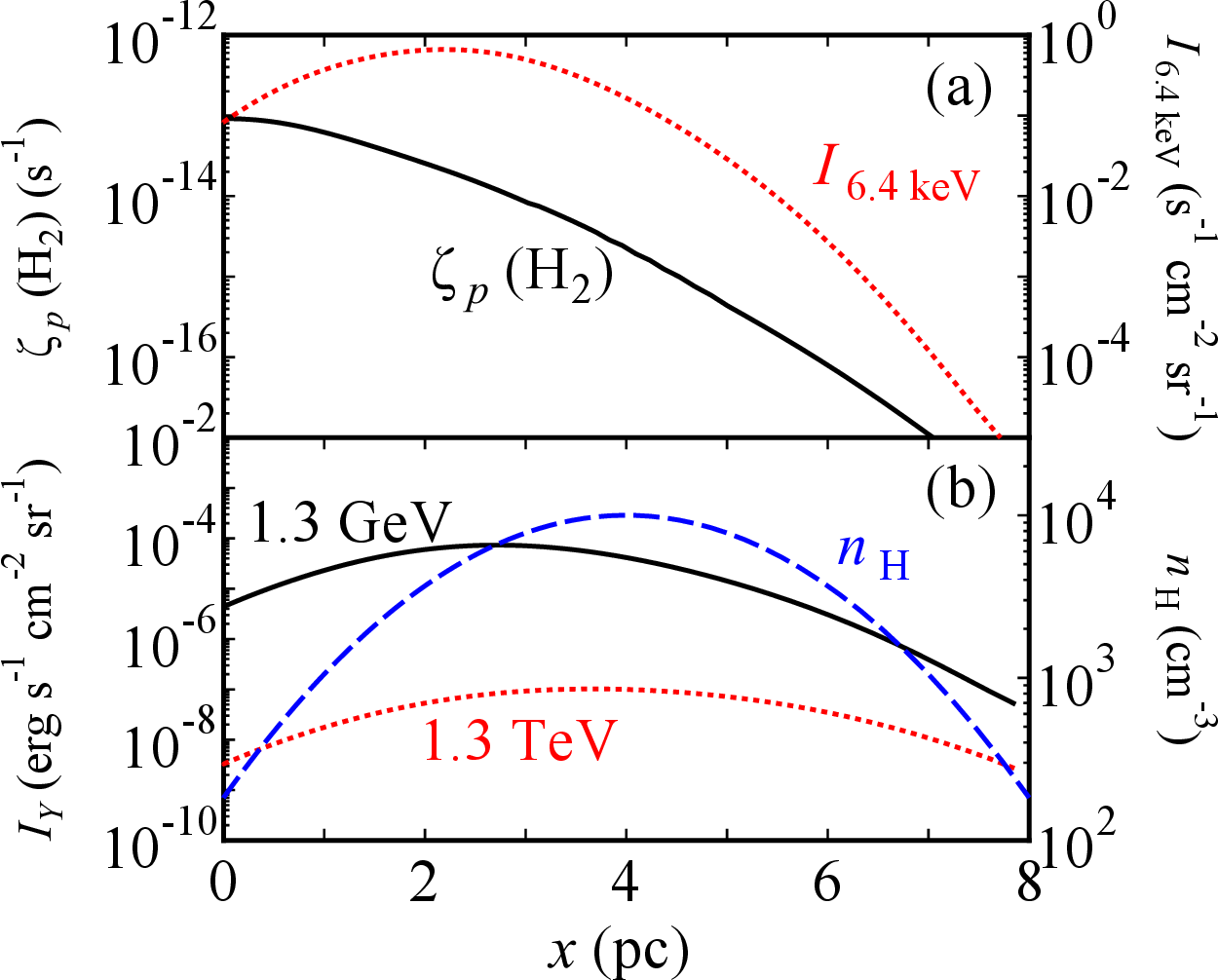} \caption{Same as Figure~\ref{fig:I_d} but for $t=1\times 10^4$~yr. \label{fig:I_d2}}
\end{figure}

\begin{figure}[t]
\plotone{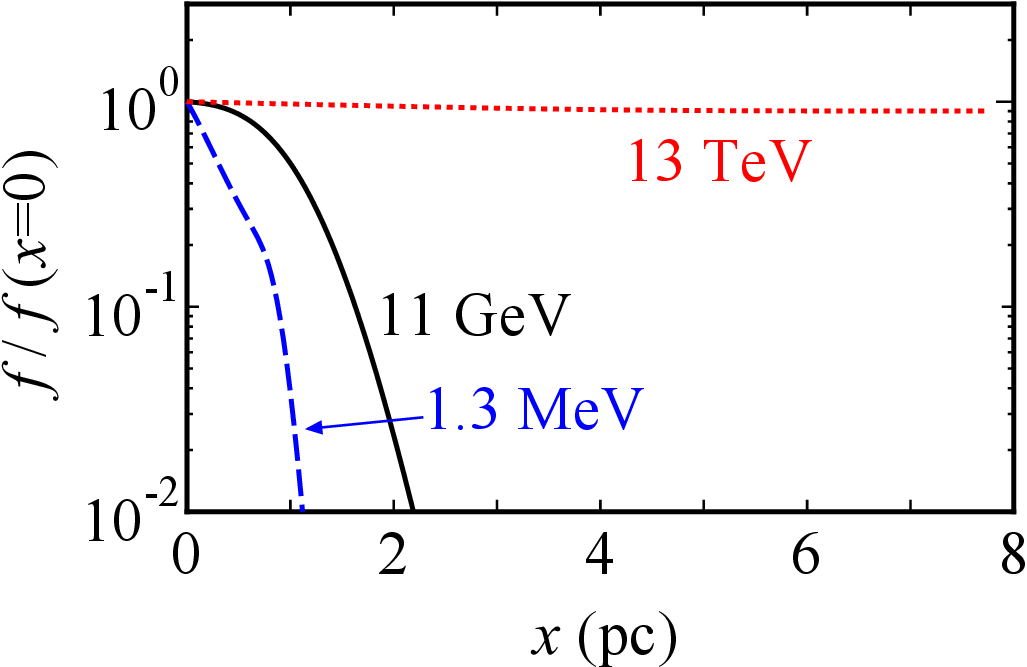} \caption{Same as Figure~\ref{fig:dist_d} ($t=3000$~yr)
but in the case where the index of the diffusion coefficient is
$\delta=1$. \label{fig:dist_d1}}
\end{figure}

\begin{figure}[t]
\vspace{2mm}
\plotone{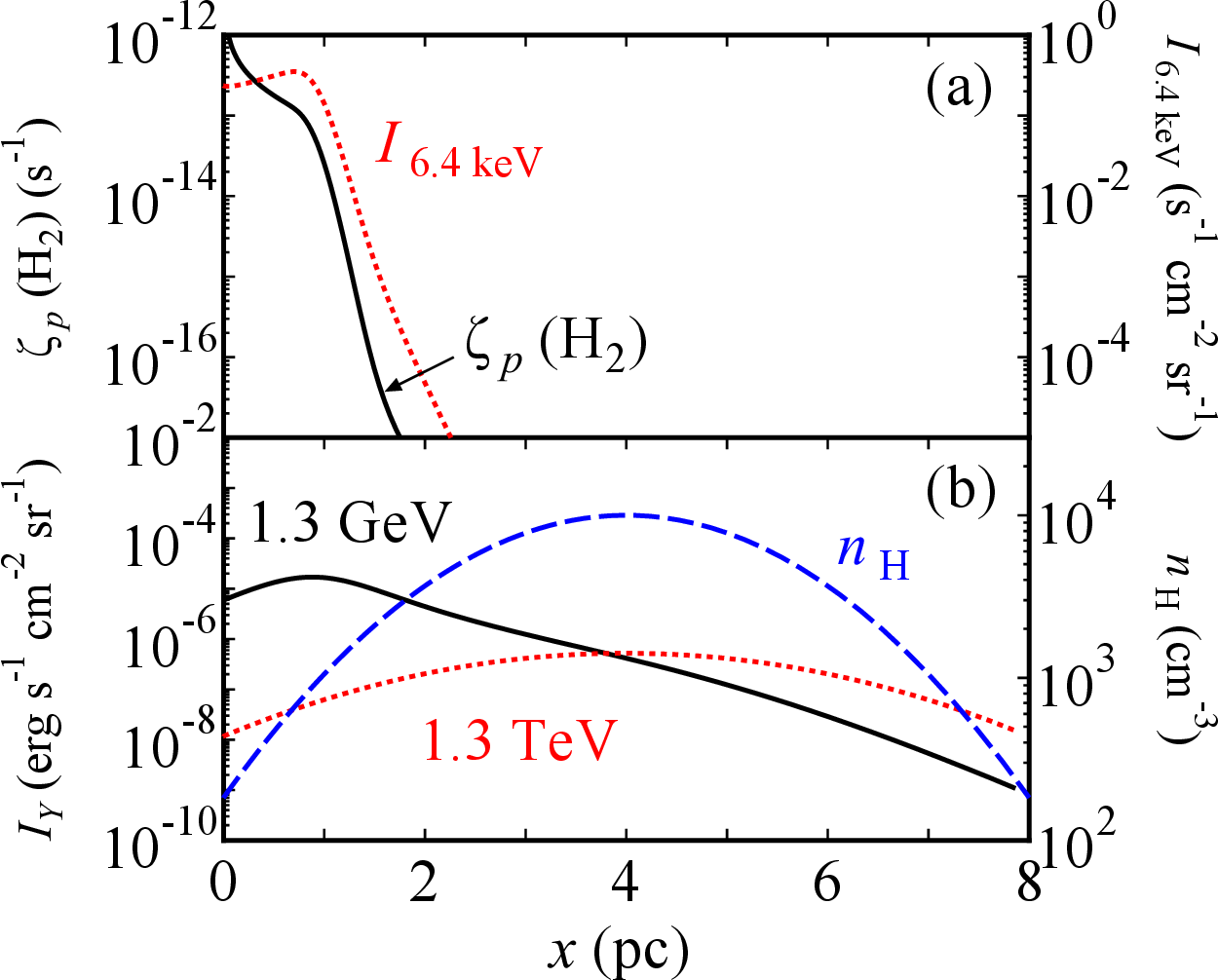} \caption{Same as Figure~\ref{fig:I_d} ($t=3000$~yr) but
in the case where the index of the diffusion coefficient is
$\delta=1$. \label{fig:I_d1}}
\end{figure}

If the line of sight is parallel to the $x$-axis (Figure~\ref{fig:mol}),
the flux of the observed 6.4~keV line is expected to be the
volume-weighted summation along the $x$-axis because the 6.4~keV line is
optically thin. In the case of Figure~\ref{fig:I_d}a, the flux along the
$x$-axis is $\langle I_{{\rm 6.4keV}}\rangle = 5.1\times 10^{-2}\rm\:
photons\: s^{-1}\: cm^{-2}\: sr^{-1}$, which is consistent with the
value observed for W~28 ($0.10^{+0.05}_{-0.05}\rm\: photons\: s^{-1}\:
cm^{-2}\: sr^{-1}$; \citealt{2018ApJ...854...87N}). If it is weighted by
$n_{\rm H}$, the ionization rate averaged along the $x$-axis is
$\langle\zeta_p({\rm H}_2)\rangle= 3.8\times 10^{-15}\rm\: s^{-1}$,
which is roughly consistent with the ones for W~28 obtained by
\citet{2014A&A...568A..50V} using DCO$^+/$HCO$^+$ abundance
ratios. However, we note that if a molecular cloud is not uniform along
the line of sight, the comparison of $\langle\zeta_p({\rm H}_2)\rangle$
with the observed ionization rate is not trivial, because the observed
rate also depends on the temperature structure of the cloud and the
abundances of molecules \citep[see
also][]{2009A&A...501..619P,2010ApJ...724.1357I,2011ApJ...740L...4C}.

Figures~\ref{fig:dist_d2} and~\ref{fig:I_d2} show the results for
$t=1\times 10^4$~yr. Compared with Figures~\ref{fig:dist_d}
and~\ref{fig:I_d} ($t=3000$~yr), CRs enter further inside the cloud and
gamma rays are produced there. While the peak of $\zeta_p({\rm H}_2)$ is
at $x=0$, that of TeV gamma rays is at $x\sim 4$~pc, where $n_{\rm H}$
reaches the maximum. The peaks of $I_{{\rm 6.4keV}}$ and GeV gamma rays
are between them.

Figures~\ref{fig:dist_d1} and~\ref{fig:I_d1} show the results
when the index of the diffusion coefficient $D(p)$ is $\delta=1$
(equation~(\ref{eq:Dp})); other parameters are the same as those of the
model of $\delta=1/3$. We consider this case because there are
observations of CR transport in the heliosphere that suggest
$D(p)\propto p$ at low energies
(\citealt{1994ApJ...420..294B,2010ApJ...719.1497S} and references
therein). Compared with the case of $\delta=1/3$, $D(p)$ is much smaller
(larger) at $pc\ll 10$~GeV ($pc\gg 10$~GeV). Thus, the lower-energy CRs
cannot penetrate deep inside the cloud, while the higher-energy CRs can
prevail in the cloud (Figure~\ref{fig:dist_d1}). This explains why the
tendencies shown in Figure~\ref{fig:I_d} are strengthened in
Figure~\ref{fig:I_d1} in which large $\zeta_p({\rm H}_2)$ and $I_{{\rm
6.4keV}}$ regions are more confined at $x\sim 0$, while TeV gamma-ray
emissions come from the entire cloud.

\subsection{Free-streaming Case}
\label{sec:ff}

\begin{figure}[t]
\plotone{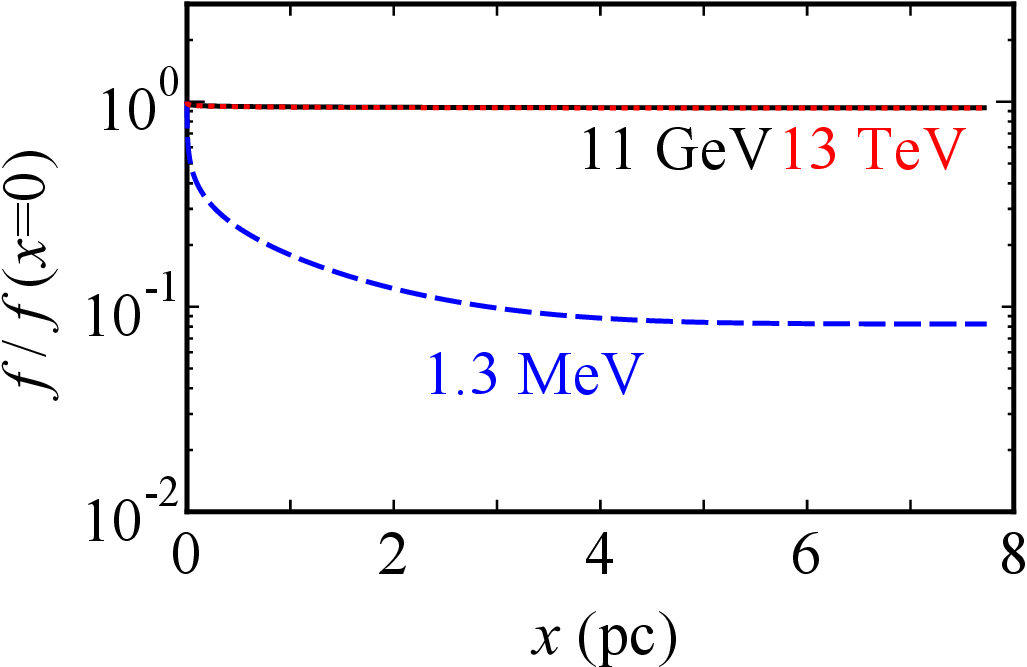} \caption{Same as Figure~\ref{fig:dist_d} ($t=3000$~yr)
but for the free-streaming case. \label{fig:dist_f}}
\end{figure}

\begin{figure}[t]
\vspace{2mm}
\plotone{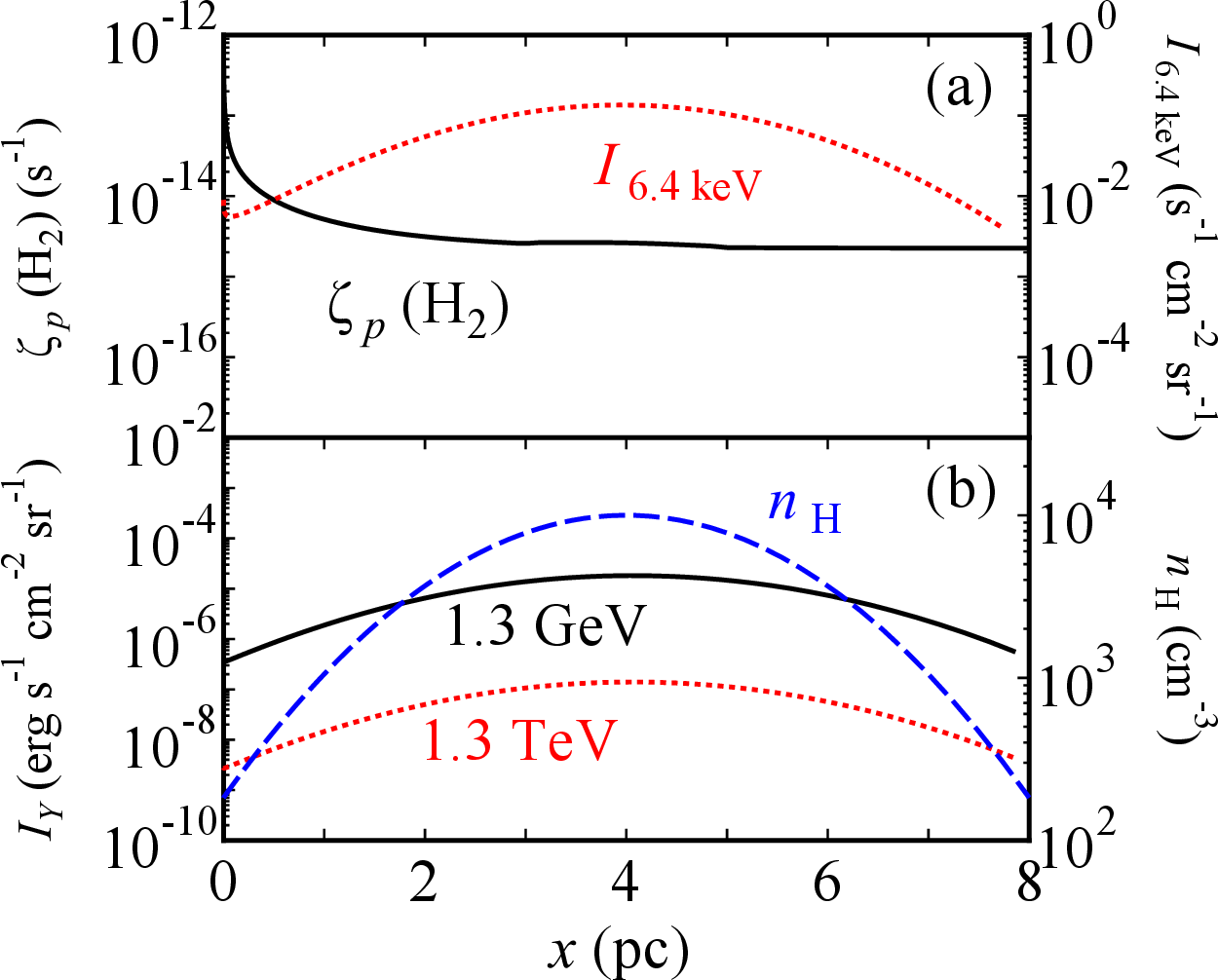} \caption{Same as Figure~\ref{fig:I_d} 
($t=3000$~yr) but for the free-streaming case. \label{fig:I_f}}
\end{figure}

Figures~\ref{fig:dist_f} and~\ref{fig:I_f} show the results for the
free-streaming case. While most parameters are the same as those for the
diffusive case, we need to reduce the total CR energy to $E_{\rm
CR,tot}=5\times 10^{48}$~erg (or increase $V_{\rm SNR}$ by 20 times) and
adopt $\alpha=2.8$ (equation~(\ref{eq:f0})) to reproduce the observed
total gamma-ray fluxes. This is because in this model both GeV and TeV
CRs freely penetrate into the cloud without cooling
(Figure~\ref{fig:dist_f}) and effectively produce gamma rays through
interaction with high-density gas at $x\sim 4$~pc
(Figure~\ref{fig:I_f}b).  The total gamma-ray fluxes are
$F_\gamma=1.3\times 10^{-10}\rm\: erg\: cm^{-2}\: s^{-1}$ at
$E_\gamma=1.3$~GeV, and $F_\gamma=1.0\times 10^{-12}\rm\: erg\:
cm^{-2}\: s^{-1}$ at $E_\gamma=1.3$~TeV. Since GeV-TeV CRs are uniformly
distributed in the cloud (Figure~\ref{fig:dist_f}), the gamma-ray
profiles simply reflect the gas density $n_{\rm H}$
(Figure~\ref{fig:I_f}b).

Contrary to the high-energy CRs, lower-energy CRs ($E\lesssim 10$~MeV)
cool before they traverse the cloud and reach $x=x_{\rm max}$. Thus,
their number density decreases as $x$ increases (blue dashed line in
Figure~\ref{fig:dist_f}) and the ionization rate $\zeta_p({\rm H}_2)$
simply reflects that profile when the line of sight ($\ell$) is parallel
to the $y$-axis (Figure~\ref{fig:I_f}a). On the other hand, the energy
of the CRs that most contribute to the 6.4~keV line is $E\sim 10$~MeV
(Figure~5 in \citealt{2012A&A...546A..88T}). These CRs are marginally
immune to the rapid cooling and are distributed almost uniformly in the
cloud. Thus, the profile of $I_{{\rm 6.4keV}}$ follows that of $n_{\rm
H}$ (equation~(\ref{eq:line formula})), which is the same as the
gamma-ray profiles (Figure~\ref{fig:I_f}). This presents a contrast to
the the diffusive case (section~\ref{sec:rdiff}).  When the
line of sight is parallel to the $x$-axis, the mass-averaged ionization
rate is $\langle\zeta_p({\rm H}_2)\rangle= 7.5\times 10^{-15}\rm\:
s^{-1}$, and the 6.4~keV line flux is $\langle I_{{\rm 6.4keV}}\rangle =
0.22\rm\; photons\; s^{-1}\: cm^{-2}\: sr^{-1}$. The results for
$t=1\times 10^4$~yr are almost the same as those for $t=3000$~yr.

\section{Discussion}
\label{sec:disc}

We have obtained the profiles of the ionization rate $\zeta_p({\rm
H}_2)$, the 6.4~keV line intensity $I_{{\rm 6.4keV}}$, and the gamma-ray
surface brightness $I_\gamma$ for a molecular cloud exposed to CRs. We
found that for both the diffusive and free-streaming CR propagation
cases, large $\zeta_p({\rm H}_2)$ regions are limited to the edge of the
cloud due to the rapid cooling of low-energy CRs that are responsible
for the ionization. In both cases, TeV gamma rays come from the dense
core of the cloud because high-energy CRs associated with the radiation
can penetrate deep into the cloud. In the diffusive case, $I_{{\rm
6.4keV}}$ has a profile similar to that of $\zeta_p({\rm H}_2)$,
although the peak is located slightly inside the cloud. The GeV
gamma-ray profile has a shape that is intermediate between $I_{{\rm
6.4keV}}$ and TeV gamma-ray profiles. In the free-streaming case, both
$I_{{\rm 6.4keV}}$ and GeV gamma rays have profiles similar to that of
TeV gamma rays. This is because CRs with $E\gtrsim 10$~MeV are uniformly
distributed in the cloud with being less affected by cooling and their
profiles reflect the gas density profile of the cloud.

While many SNRs have been observed in gamma rays, the number of SNRs for
which ionization rates, 6.4~keV line and gamma-ray fluxes are all
obtained is limited. The exceptions are W~28, IC~443, and W~51C. Since
our model is rather simple, we here qualitatively compare our
predictions with the observations. For W~28, ionization rates have been
obtained through DCO$^+$/HCO$^+$ abundance ratios
\citep{2014A&A...568A..50V} and 6.4~keV line fluxes have been measured
with Suzaku \citep{2018ApJ...854...87N,2018PASJ...70...35O}.
\citet{2018ApJ...854...87N} indicated that the 6.4~keV line emitting
region is shifted from the center of the northeastern molecular cloud
and the TeV gamma-ray emitting region \citep{2008A&A...481..401A}, while
\citet{2018PASJ...70...35O} detected the 6.4~keV line in the gamma-ray
emitting region. The former and the latter favor the diffusive case
(section~\ref{sec:rdiff}) and the free-streaming case
(section~\ref{sec:ff}), respectively.  \citet{2018ApJ...854...87N}
analyzed X-ray spectra for regions selected based on an X-ray image in a
narrow band including the 6.4~keV line, while
\citet{2018PASJ...70...35O} performed spectral analysis for a few
specific regions irrespective of the narrow band image. The apparent
discrepancy of the position of the line emitting region may indicate
some uncertainty in current X-ray observations. High ionization rates
are discovered at the tip of the cloud \citep[Figure~1
in][]{2020A&A...635A..40P}.  Unfortunately, ionization rates are studied
only for limited areas and they are not obtained for the regions where
the 6.4~keV line is detected by \citet{2018ApJ...854...87N}. If the
diffusive case is correct, the ionization rates should be high there.

For W~28, we examine the possibility that the 6.4~keV line is
created by X-rays rather than CRs. Figure~5 in
\citet{2018PASJ...70...35O} shows that the X-ray radiation at the energy
of $E_X > 7$~keV, which is responsible for the 6.4~keV line, is
dominated by background emissions. The intensity of the 6.4~keV line
created by the emissions is estimated as
\begin{equation}
 I_{\rm 6.4keV} = \epsilon\int_{E_{\rm Ked}}^{\infty} 
[1-\exp{(-N_{\rm H} Z_{\rm Fe}\sigma_{\rm Fe}(E_X))}] I_b(E_X)dE_X\:,
\end{equation}
where $\epsilon=0.34$ \citep{1972RvMP...44..716B} is the iron
fluorescence yield, $E_{\rm Ked}=7.1$~keV is the energy of the iron
K-edge, $N_{\rm H}$ is the hydrogen column density of the molecular
cloud, $Z_{\rm Fe}$ is the iron abundance, $\sigma_{\rm Fe}(E_X)$ is the
cross-section of the iron photoionization, and $I_b(E_X)$ is the flux of
the X-ray background. Here, we adopt $\sigma_{\rm Fe}(E_X)\approx
6.0\times 10^{-18} E_X^{-2.6}\rm\: cm^2$
\citep{1982ADNDT..27....1H}. The background flux is estimated from that
in the reference region studied by \citet{2018ApJ...854...87N} and it is
represented by $I_b(E_X) \approx 1.2\times 10^{-5}\: E^{-2.3}\rm\:
photons\: keV^{-1}\: cm^{-2}\: s^{-1}\: arcmin^{-2}$. If we assume that
$N_H=3\times 10^{22}\rm\: cm^{-2}$ and $Z_{\rm Fe}=3\times 10^{-5}$
(solar abundance; \citealt{2003ApJ...591.1220L}), we obtain $I_{\rm
6.4keV}=3.0\times 10^{-9}\rm\: photons\: cm^{-2}\: s^{-1}\:
arcmin^{-2}$, which is 1 order of magnitude smaller than those
observationally obtained by \citet{2018ApJ...854...87N} and
\citet{2018PASJ...70...35O}. If the value of $N_{\rm H}$ is smaller
around the cloud edge (see equation~(\ref{eq:nH})), $I_{\rm 6.4keV}$
becomes even smaller. Thus, we conclude that the contribution of X-rays
to the observed 6.4~keV line intensity is minor.

For IC~443, one of the regions that are emitting the 6.4~keV line
\citep[Reg~2 in Figure~1 of][]{2019PASJ...71..115N} is close to the
centroids of GeV--TeV gamma-ray sources
\citep{2007ApJ...664L..87A,2009ApJ...698L.133A,2010ApJ...712..459A},
which favors the free-streaming case (Figure~\ref{fig:I_f}). Another
region from which the 6.4~keV line is detected \citep[Reg~1 in Figure~1
of][]{2019PASJ...71..115N} is shifted from the gamma-ray
centroids. Ionization rates have been measured for several points around
the SNR \citep{2010ApJ...724.1357I}. Two of the points (ALS~8828 and
HD~254577) show high ionization rates. We found that ALS~8828 is very
close to Reg~1. Thus, Reg~1 may be described by the diffusive case
(Figure~\ref{fig:I_d}). HD~254577 is not associated with
gamma rays. There is no report about the 6.4~keV line at HD~254577 and
there is no measurement of ionization rates in Reg~2.

For W~51C, it has been indicated that a high ionization point coincides
with a gamma-ray source (\citealt{2011ApJ...740L...4C}; see also
\citealt{2009ApJ...706L...1A,2012A&A...541A..13A}). Suzaku X-ray
observations also suggest that the 6.4~keV line emission comes from the
same region (Shimaguchi et al. in preparation). However, the relatively
small angular size of the SNR does not allow us to discuss their relative
positions in detail.

We note that in the above discussions we implicitly assumed that the
line of sight is perpendicular to the direction of CR diffusion or
stream ($x$-axis in Figure~\ref{fig:mol}). If the line of sight is
parallel to the $x$-axis, high ionization rates could be measured where
both the 6.4~keV line and gamma-ray emissions are detected, regardless
of the diffusive and free-streaming cases. Reg~2 in IC~443 could be such
an example because it is apparently located well inside the the SNR
\citep{2019PASJ...71..115N} and three-dimensionally it could be behind
or in front of the SNR. Statistical studies with more samples may be
required to investigate the three-dimensional effect. In
summary, so far there is no strong evidence that either of the diffusive
or free-streaming intrusion is dominant considering observational
uncertainties.

In the near future, the advent of XRISM and Athena will make
it possible to measure the spectral profile of the 6.4~keV line in more
detail. Recently, \citet{2020PASJ...72L...7O} indicated that the iron
line substructures generated by the multiple ionization process can be
direct evidence that the line is produced by CRs rather than
X-rays. Those satellites will be able to resolve the substructures. For
ionization rates, it is desirable to make intensive observations that
cover wide fields around SNRs using radio telescopes such as Nobeyama
45~m and ALMA.

\section{Conclusion}
\label{sec:conc}

Using an 1D model, we have studied the spatial correlation among the
ionization rates of dense gas, the 6.4~keV line intensity, and the
gamma-ray emissions from a molecular cloud illuminated by CRs
accelerated at an adjacent SNR. We found that the profiles of the three
observables depend on how CRs intrude the cloud and on the internal
structure of the cloud. If the CRs diffusively enter the cloud and the
diffusion coefficient is relatively small, the 6.4~keV line should be
detected in the cloud outskirts where ionization rates are high.  This
is because the 6.4~keV line and the ionization are attributed to CRs
with energies of $\lesssim 10$~MeV. Those CRs cool before they penetrate
the cloud.  On the other hand, if CRs freely move in the cloud, the
6.4~keV line profiles should be similar to gamma-ray profiles. The
energies of the CRs associated with the 6.4~keV line ($\sim 10$~MeV) are
higher than those associated with the ionization ($\sim
0.01$~MeV). Thus, the former can traverse the cloud before cooling in
the same way as gamma-ray emitting CRs with higher energies. Both the
6.4~keV line flux and gamma-ray profiles follow the gas profiles of the
cloud.  We compared the results with observations and argued
whether the diffusive or the free-streaming intrusion is
realized. However, we could not draw an unambiguous conclusion mostly
because of observational uncertainties.

\acknowledgments

We would like to thank the anonymous referee for useful comments. This
work was supported by MEXT KAKENHI No.18K03647, 20H00181 (YF), 20K14491
(KKN), and 19K14758, 19H05075 (HS).

\bibliographystyle{aasjournal}

\end{document}